\documentstyle[12pt]{article}
\begin{document}
\def\be{\begin{equation}}
\def\ee{\end{equation}}
\def\bc{\begin{center}}
\def\ec{\end{center}}
\def\beq{\begin{eqnarray}}
\def\eeq{\end{eqnarray}}
\def\beqd{\begin{eqnarray*}}
\def\eeqd{\end{eqnarray*}}
\def\nin{\noindent}
\def\lra{$\leftrightarrow$ }

\def\etaten{$\eta_{10}$}
\def\he{$^4$He}
\def\dhe{D$+^3$He/H}
\def\rhoe{$\rho_e$}
\def\Pe{$P_e$}
\def\drhoedt{$d\rho_e/dT_{\gamma}$}
\def\tnu{T_{\nu}}
\def\rhonu{\rho_{\nu}}
\def\nnu{N_\nu}
\def\dnnu{\Delta \nnu}
\def\tg{T_{\gamma}}

\rightline{{\bf CWRU-P22-96}}
\rightline{December 1996}
\baselineskip=16pt
\vskip 0.5in
\begin{center}
\bf\large {Comment on ``Constraints on the strength of primordial
B-fields from big bang nucleosynthesis reexamined''}
\end{center}
\vskip0.2in
\begin{center}
Peter J. Kernan, Glenn D. Starkman
and
Tanmay Vachaspati
\vskip .1in
 {\small\it Department of Physics\\
Case Western Reserve University\\ 
10900 Euclid Ave., Cleveland, OH 44106-7079}
\vskip 0.4in
\end{center}
\centerline{{\bf Abstract}}
\noindent
Recently Cheng, Olinto, Schramm and Truran~\cite{COST} 
reexamined the constraints on the strength of primordial
B-fields from big bang nucleosynthesis (BBN).  
Their  bottom line agreed with that of an earlier recent paper
on the subject~\cite{KSV}, both in its final limit on the
B-field during BBN, and in its conclusion that for allowed values
of the B-field, the dominant 
factor for BBN is the increased expansion rate at a given 
temperature
caused by the energy density of the magnetic field,
$B^2/8\pi$. 
However, their conclusion that weak interaction rates increased
with increasing B-field at these low field values
contradicted the earlier results of \cite{KSV}.
In this comment we point out that the Taylor series expansion 
of the weak interaction rate  about $B=0$ used in \cite{COST} 
is not well-defined, while the
Euler-McLaurin expansion of \cite{KSV} is well-behaved and reliable.
Using the Euler-McLaurin expansion we find that 
the weak interaction rates decrease rather than increase 
with increasing B-field at small values of the B-field.

\vfill
\eject

%\section{Introduction}

As discussed by Cheng, Olinto, Schramm and Truran~\cite{COST}
{\lq\lq}BBN provides a unique quantitative window for processes
occurring in the early universe between temperatures of $1-0.01$MeV.
A primeval magnetic field existing during this 
period would have three major  effects on BBN" one of which
is that it would alter the weak interaction rates.
Equations (2.4-2.6) of  ~\cite{COST} give the reaction 
rates for $ne^+\to p{\bar\nu}_e$,
$n\nu_e\to pe^-$ and 
$n\to pe^-\nu_e$ respectively.
For example the rate for $n\nu\to pe^-$ is
\beq
\lambda_{n\nu\to pe^-}&={g_V^2(1+3\alpha^2)m_e^5\gamma_iT_\nu^2
			\over 4\pi^3T_i^2}\times
\phantom{junkjunkjunkjunkjunkjunk}
\label{theirs}
\\
&\times\bigl[\sum_{n_s=0}^{\infty}[2-\delta_{n_s0}]
\int_{\sqrt{1+2\gamma n_s}+\kappa}^{\infty}
{d\epsilon(\epsilon-\kappa)(\epsilon-q)^2 
\over\sqrt{(\epsilon-\kappa)^2-(1+2\gamma n_s)}}
{e^{\epsilon Z_e + \phi_e}\over 1 + e^{\epsilon Z_e + \phi_e}}
{1\over 1 + e^{(\epsilon-q) Z_\nu + \phi_\nu}}
\nonumber
\\
-&\sum_{n_s=0}^{n_{s max}+1}[2-\delta_{n_s0}]
\int_{\sqrt{1+2\gamma n_s}+\kappa}^q
{d\epsilon(\epsilon-\kappa)(\epsilon-q)^2 
\over\sqrt{(\epsilon-\kappa)^2-(1+2\gamma n_s)}}
{e^{\epsilon Z_e + \phi_e}\over 1 + e^{\epsilon Z_e + \phi_e}}
{1\over 1 + e^{(\epsilon-q) Z_\nu - \phi_\nu}}\bigr]
\nonumber
\eeq
where $\alpha=g_V/g_A$,$\gamma=B/B_c$ (with $B_c=m_e^2/e$),
$\gamma_i = B(T=1MeV)/B_c$,$n_s=n+{1\over 2}-s_z$ (with $n$ the
principal quantum number of the Landau level and $s_z=\pm{1\over2}$
the spins, $\kappa$ is the anomalous magnetic moment term 
for an electron in the ground state ($n=0$, $s_z=1/2$), 
$\epsilon=E_e/m_e$, $q=(m_n-m_p)/m_e$, $Z_e=m_e/T_e$, 
$Z_\nu=m_\nu/T_\nu$,
$\phi_e=\mu_e/T_e$, $\phi_\mu=\mu_\mu/T_\mu$, $m_i$ are the rest
masses of species $i$, $T_i$ are their temperatures 
and $\mu_i$ their chemical potentials. 
$n_{s max}$ is the largest integer in $[(q-\kappa)^2-1]/2\gamma$.
In the absence of chemical potentials and ignoring the anomalous 
electron magnetic moment both of which are lower order effects
at the magnetic fields which ultimately prove to be of interest
(unless there is a non-standard neutrino chemical potential)
this can be rewritten as (equation (22) of \cite{KSV}):
\be
\lambda_{n\nu\to pe^-} = {G_F^2 \tg^2 (g_V^2 + 3g_A^2) 
\over (2\pi)^3}
z  \sum_{n=0}^{\infty} (2-\delta_{n0})
\int_{0}^{\infty} dp_z E_{\nu}^2 g(E_e/\tg)f(E_{\nu}/\tnu)
\label{ours}
\ee
where we have rewritten the integral over the electron energy
in terms of an integral over the electron  momentum parallel to the
field, $p_z$.  Here  $G_F$ is the Fermi constant, $\tg=T_e$ is
the photon temperature, $f(E_{\nu}/\tnu)$ is the Fermi-Dirac
distribution, and $g=1-f$ is the Fermi blocking factor,
$z\equiv 2eB/T_\gamma^2$ 
(so that also $z=2 e \gamma_i {T_\nu^2\over \tg^2}
{m_e^2\over e T_i^2}$).
The electron energy and momentum are related by
\be
E_e = (p_z^2 + m_e^2 + 2eBn)^{1/2}\ ,
\ee
while energy conservation gives:
\be
E_{\nu} = E_e - (m_n-m_p) \equiv E_e - \Delta,
\ee
and
$f(E_{\nu}/\tnu)$ is taken to be zero if $E_\nu<0$.
We notice that $\lambda$ is of the form $z h(z)$,
where $h(z)$ is a complicated function of $z$, and so 
\be
{d\lambda\over dz}  = h(z) + z {d h(z)\over d z}
\ee
We wish to examine the weak field limit $z\to 0$ in
order to understand whether magnetic fields speed-up or
slow-down weak interactions.
The temptation (see \cite{COST}) 
is to set the second term to zero, and notice
that $h(z)$ is the sum of positive terms, and thus conclude that
${d\lambda\over dz}  > 0$.
The problem is that 
\be
\lim_{z\to0}h(z) = \infty
\ee 
We can see this explicitly by setting $B=0$ ({\it i.e.}
$\gamma = 0 =\gamma_i$)  in (\ref{theirs})
and noticing that (with $\kappa=\phi_i=0$) the
argument of the integral is finite, positive and
independent of $n_s$, hence the sum from $n_s=0,...,\infty$
is badly divergent.
In fact $h(z)\propto 1/z$ at small $z$ since $\lambda (z)
=zh(z)$ should go to the interaction rate in the absence
of a magnetic field as we let $z \rightarrow 0$.

The Taylor series of $\lambda$ about $B=0$ is therefore ill-defined.  
However, we can use the Euler-McLaurin expansion
for $\lambda$ (\cite{KSV} equation (23)):
\beq
\lambda_{n\nu\to pe^-} &=& {G_F^2 \tg^2 (g_V^2 + 3g_A^2) 
\over (2\pi)^3}
\bigl\{ 2\int_0^\infty dx \int_{p_{min}(x)}^{\infty} 
dp_z G_-(E_e(p_z,x)) 
\label{lambdannu}\\
&-& {z^2 \tg^2 \over 12}  \int_{p_{min}(0)}^{\infty}
dp_z \left[ {1\over E_e}{d G_-(E_e)\over dE_e}\right]_{x=0}
\bigr\}  + {\cal O}(z^3)
\nonumber
\eeq
where
\be
E_e(p_z,x) = (p_z^2 + m_e^2 + x)^{1/2},
\nonumber
\ee
\be
p_{min}(x) = (\Delta^2 - m_e^2 - x \tg^2)^{1/2}
\nonumber
\ee
and
\be
G_-(E_e) = (E_e - \Delta)^2 {1\over 1+ e^{(E_e -\Delta)/T_\nu}}
                            {1\over 1+ e^{-E_e/\tg}}.
\ee
Thus
\beq
\left[ {1\over E_e}{d G_-(E_e)\over dE_e}\right] &=&
{E_e - \Delta\over E_e} {1\over 1+ e^{(E_e -\Delta)/T_\nu}}
                 {1\over 1+ e^{-E_e/\tg}}
\label{useful}\\
&\times&\left[2 - {E_e - \Delta\over T_\nu}
         {1\over 1+ e^{-(E_e -\Delta)/T_\nu}}
        + {E_e\over \tg} {1\over 1+ e^{E_e/\tg}}\right] 
\nonumber
\eeq

The terms of the series in (7) are perfectly finite 
in the limit $z\to0$.
The zeroth order term recovers exactly the usual $B=0$ result.
Interestingly, we find by explicit calculation that the  
first order term in $B$ actually vanishes,
in apparent contradiction to the positivity of the first order
term in the Taylor series. This is of course because the Taylor
expansion is not well defined; 
every term of the Taylor series contributes  at arbitrarily  small
B-field. 

The sign of the $z^2$ term in $\lambda$ can be seen by inspection of
(\ref{useful}).
We see that only the second term 
in the expression in square brackets is negative 
and that for $x=0$, 
by the time $E_e$ is large enough for this term 
to dominate over the other two terms in the square brackets,
the prefactor is exponentially suppressed.
The coefficient of the $z^2$ term of (\ref{lambdannu}) 
is therefore negative,
and the B-field decreases this rate.
Similar arguments can be applied to the other $2\to 2$ rates.
Though the conclusions are not always so clear analytically,
numerical studies ~\cite{KSV} show
a decrease in all the 2-body rates for $B > 0$,
in contradiction to the 
conclusions of ~\cite{COST}.
Fortunately as both ~\cite{COST} and ~\cite{KSV} agree,
this effect is subdominant; the dominant effect of
the B-field on nucleosynthesis is simply the $B^2$ contribution 
to the energy density, which increases the expansion rate.

\end{document}